# Prediction and observation of topological modes in fractal nonlinear optics

Boris A. Malomed

Department of Physical Electronics, School of Electrical Engineering, Faculty of Engineering, and Center for Light-Matter Interaction, Tel Aviv University, Tel Aviv 69978, Israel

## Abstract

This item from the News & Views category, to be published in *Light: Science & Applications*, aims to provide a summary of theoretical and experimental results recently published in Ref. [24], which demonstrate the creation of corner modes in nonlinear optical waveguides of the higher-order topological-insulator (HOTI) type. Actually, these are second-order HOTIs, in which the transverse dimension of the topologically protected edge modes is smaller than the bulk dimension (it is 2, in the case of optical waveguide) by 2, implying zero dimension of the protected modes, that are actually realized as corner or defect ones. Work [24] reports prediction and creation of various forms of the corner modes in a HOTI with a fractal transverse structure, represented by the Sierpiński gasket (SG). The self-focusing nonlinearity of the waveguide's material transforms the corner modes into corner solitons, almost all of which are stable. The solitons may be attached to external or internal corners created by the underlying SG. This N&V item offers an overview of these new findings reported in Ref. [24] and other recent works, and a brief discussion of directions for the further work on this topic.

It is well established that the propagation through specially designed waveguides may impart various topological structures to light waves. The variety of topological states carried by light is greatly enhanced by intrinsic nonlinearity of the optical medium, which, in most cases, amounts to the Kerr (alias $\chi^{(3)}$) self-focusing, represented by cubic terms in the corresponding propagation equations. The simplest example is provided by the two-dimensional (2D) photonic crystal with the transverse square-lattice structure: this waveguide, built in a self-focusing material, readily supports the stable propagation of self-trapped optical modes with *embedded vorticity*, i.e., *vortex solitons* [1,2]. The integer value of the vorticity plays the role of the respective topological charge. These vortex solitons are arranged as multipeak patterns, the vorticity being defined as the respective *winding number*, i.e., the total phase gain along a trajectory surrounding the vortex' pivot, divided by $2\pi$. It is relevant to mention that similar stable vortex solitons may be maintained not only by spatially periodic underlying lattices but also by quasiperiodic ones (i.e., *photonic quasicrystals*) [3].

A related vast topic in studies on linear and nonlinear optics is *emulation*, by means of light fields, of various phenomena which are known in a much more complex form in solid-state physics, a popular example being the creation of photonic counterparts of graphene [4-7]. In particular, much interest was drawn to the studies of edge modes [4,5] and spin-orbit-coupling [7] in the photonic graphene. These setups are also based on light propagation in lattice structures.

A class of solid-state settings which has drawn a great deal of interest in the course of past 20 years comprises topological insulators (TIs, alias quantum spin Hall insulators) [8,9], see also reviews [10,11]. TI is a crystalline material bounded by surfaces, which is an insulator in the bulk, with a finite gap in its excitation spectrum, while the surface (edge) states feature a topologically protected gapless (conducting) spectrum. The topological protection implies that the surface conductivity persists in the presence of defects and irregularities. Following the discovery of TIs in solid-state physics, their photonic counterparts have been created, also demonstrating the topological protection of the surface states [12,13] (it is relevant to compare it to the recently demonstrated topological protection of optical skyrmions [14]).

Further work on the topic of TIs has led to the discovery of higher-order TIs (HOTIs). Their definition states that the TI of order $m$ realized in the bulk medium of dimension $D$ supports the topologically protected conductivity on a surface of dimension $D - m$ [15-17]. In terms of the realization in photonics, which has $D = 2$, this implies that, in addition to the usual (first-order) TI, with $m = 1$, one can consider the second-order HOTI, with $m = 2$. The respective zero value of the surface's dimension implies that the topologically protected conducting states may exist as *corner modes*, attached to junctions of orthogonal surfaces, or ones attached to local defects [18-20]. Furthermore, it has been demonstrated, theoretically and experimentally, that the self-focusing nonlinearity transforms such corner modes into similarly structured spatial solitons [21,22]. In this connection, it is relevant to mention that 2D solitons in the uniform medium with the cubic self-focusing are unstable, due to the occurrence of the *critical collapse* in the same setting [23]. However, lattice potentials, periodic [1,2] and aperiodic [3] ones alike, can readily stabilize both fundamental and vortex solitons in this case.

The new work [24], published in journal Light: Science & Applications, reports prediction and experimental realization of a great expansion of the variety of HOTI states in an aperiodic nonlinear photonic crystal, in which the underlying lattice structure is *fractal*, realized in the form of a *Sierpiński gasket* (SG). In its ideal form, SG is a carpet of infinitely downscaling equilateral triangles (the texture of this pattern can be seen below in Figs. 1 and 2; note that its aperiodicity makes it somewhat cognate to the above-mentioned photonic quasicrystals – in particular, because practically available patterns do not feature infinite downscaling of the triangular structure). The SG-shaped waveguide used in the experimental part of [24] was fabricated by means of the technique which inscribes a desirable structure in fused silica by femtosecond pulses of UV light. An essential peculiarity of the SG structure, considered in a finite domain, is the fact that it features both external corners and internal ones, formed by triangular holes in the SG (see Figs. 1 and 2). Thus, aiming to construct SG-maintained HOTIs, one may expect to find such localized modes attached to both external and internal corners of the underlying fractal lattice of a finite size. Of course, both numerical and experimental results reported in Ref. [24] were actually obtained, as mentioned above, not for the truly fractal setting, but for some approximation to it.

It is relevant to mention that a large variety of chiral corner modes in a linear TI waveguide based on the SG lattice of the *Floquet type*, i.e., one which is periodically modulated along the propagation distance, were recently created in the experiment [25]. However, effects of the nonlinearity were not addressed in that work.

The analysis developed in the theoretical part of work [24] is based on the nonlinear Schrödinger equation for amplitude $\psi$ of the optical wave, written in the scaled form:

$$i\frac{\partial \psi}{\partial z} = -\frac{1}{2}\left(\frac{\partial^2}{\partial x^2} + \frac{\partial^2}{\partial y^2}\right)\psi - R(x,y)\psi - |\psi|^2\psi, \tag{1}$$

where $z$ and $(x,y)$ are, respectively, the propagation distance and transverse coordinates, real function $R(x,y)$ represents the effective potential induced by the underlying SG fractal structure, the Laplacian operator represents the paraxial diffraction, and the sign minus in front of the cubic term corresponds to the self-focusing sign of the nonlinearity.

The SG-supported corner modes are produced by a numerical solution of the linearized version of Eq. (1), by means substitution

$$\psi(x,y;z) = \exp(ibz)\,u(x,y), \tag{2}$$

where $b$ is a real propagation constant, and stationary field $u(x,y)$ is real too. As mentioned above, such localized modes are attached to the external or internal corners of the SG lattice (actually, all corners may support localized modes attached to them). Hybrid modes, that feature several local peaks that are pinned to both external and internal corners, have been found too. If the linearized equation (1) is considered as the 2D quantum-mechanical Schrödinger equation, the corner modes are construed as bound states supported by the SG potential $R(x,y)$. In terms of the spectrum of eigenstates of the linearized equation, the localized corner modes belong to *bandgaps*, i.e., intervals of values of the propagation constant at which extended (delocalized) linear modes do not exist.

The numerical solution of full equation (1), which includes the cubic term, demonstrates the transformation of the various linear corner modes into stable *corner solitons*, which inherit the symmetry and general shape of the parent corner modes (in this connection, the solitons are usually referred to as ones *bifurcating from* the underlying localized linear modes). Soliton families are characterized by the dependence of their integral power, $P \equiv \iint u^2(x,y)dxdy$, on the propagation constant $b$ (see equation (2)). These dependences are *thresholdless*, i.e., they start from $P = 0$, at the bifurcation point (a finite threshold would mean that a soliton family starts from finite $P > 0$). Further, as well as the linear modes from which they bifurcate, the soliton families may exist solely at values of $b$ which belong to one of the bandgaps of the underlying spectrum of the linearized equation (1). A characteristic set of numerically found corner-soliton solutions, borrowed from [24], is displayed in Fig. 1.

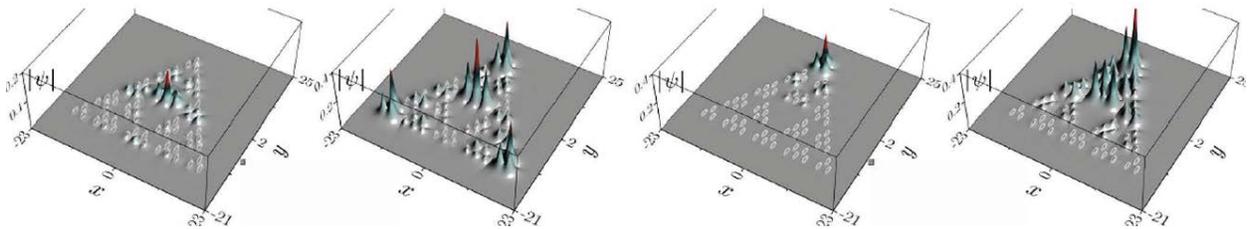

**Fig. 1**. A set of stable corner solitons produced by the numerical solution of Eq. (1) for fixed SG parameters and different values of the propagation constant $b$. It is a fragment of figure 2 from Ref. [24].

The change of the corner-solitons' shapes with the increase of $P$ makes it possible to use the power as a tool which helps to "morph" the solitons. Eventually, pumping more power into them, one pushes the corner solitons from original bandgaps, in which they bifurcated from the linear corner modes, into adjacent *Bloch bands*, where the former solitons develop nonvanishing tails, thus transforming themselves into *quasisolitons*.

The analysis of stability of the corner solitons against small perturbations, as well as direct simulations of their perturbed evolution, demonstrates that almost all of them are stable [24].

In Ref.[ [24] the experimental creation of corner solitons was performed in an experimental sample with the propagation distance 10 cm, which is sufficient for the demonstration of the self-trapping of various species of the corner solitons. A set of observed quasi-soliton profiles is displayed, for increasing values of the energy ($E$) of input pulses, i.e., for growing levels of power $P$, in Fig. 2. It is clearly seen that the multipeak corner solitons, found at low levels of $P$, demonstrate strong self-compression and reduction of the number of peaks with the increase of $P$. The above-mentioned transformation of the solitons into quasi-solitons with nonvanishing tails was not observed experimentally, as the power levels needed for the transformation would damage the material.

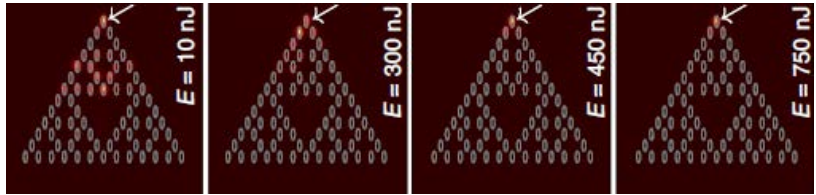

**Fig. 2**. A set of images of the quasi-soliton corner modes (occupying reddish sites of the underlying SG), as reported in Ref. [24] (see Fig. 4 in that work) at increasing values of the total power (it is proportional to energy $E$ of the input light pulse, as indicated in panels of the figure). The white arrows indicate the site at which the input pulse was shone into the sample.

**Conclusion**

The results reported in detail in Ref. [24] and outlined above are of considerable interest for fundamental studies in the field of nonlinear photonics, and may find applications – e.g., for the design of topological lasers and generation of spatial solitons with required shapes.

The analysis reported in [24] may be extended in other directions. First, it is possible to seek for bound states of broad corner solitons centered at adjacent external and/or internal corners (such composite states were not reported in Ref. [24]). Next, it would be relevant to construct corner solitons with embedded vorticity (for the theoretical analysis of this option, one should use substitution (2) with a complex stationary function $u(x, y)$, whose phase circulation may carry the vorticity [1-3], as mentioned above). Further, one can consider the TI waveguide with the *self-defocusing* nonlinearity, i.e., the opposite sign in front of the cubic term in Eq. (1). In that case, the formation of fundamental and vortex *gap-soliton* modes may be expected [26,27,3]. In particular, unlike the corner solitons which are rigidly pinned to the underlying lattice in the self-focusing medium, 2D gap solitons may demonstrate mobility in the defocusing medium [27]. Finally, it may also be interesting to create corner solitons in photonic TIs with the quadratic (*second-harmonic-generating*, alias $\chi^{(2)}$ [28]), rather than cubic, material nonlinearity.